\documentclass[showpacs,preprintnumbers,eqsecnum,twocolumn,prd]{revtex4-1}  

\usepackage{graphicx}
\usepackage{latexsym}
\usepackage{amsmath}
\usepackage{amssymb}

\def\beq{\begin{equation}}
\def\eeq{\end{equation}}

\def\bfOmega{\pmb{\Omega}}

\begin{document}

\title{Analytic determination of high-order post-Newtonian self-force contributions to gravitational spin precession}

\author{Donato \surname{Bini}$^1$}
\author{Thibault \surname{Damour}$^2$}

\affiliation{$^1$Istituto per le Applicazioni del Calcolo ``M. Picone'', CNR, I-00185 Rome, Italy\\
$^2$Institut des Hautes Etudes Scientifiques, 91440 Bures-sur-Yvette, France}

\date{\today}
\begin{abstract}
Continuing our analytic computation of  the first-order self-force contribution to the \lq\lq geodetic" spin precession frequency of a small spinning body orbiting a large (non-spinning) body we provide the exact
expressions of the  tenth and tenth-and-a-half post-Newtonian terms. We also introduce a new approach to the analytic computation of self-force regularization parameters
based on a WKB analysis of the radial and angular equations satisfied by the metric perturbations.
\end{abstract}

\maketitle

\section{Introduction}

The impending prospect of detecting gravitational-wave signals from coalescing compact binary systems motivates renewed studies of the general relativistic
dynamics of  binary systems made of {\it spinning} bodies.  It has been emphasized in  Ref.~\cite{Damour:2007nc} that a simple way of computing (to linear
order in each spin) the spin-dependent interaction terms  $H_{\rm int} = {\bfOmega}_1^{\rm SO}\cdot {\mathbf S}_1 + {\bfOmega}_2^{\rm SO}\cdot {\mathbf S}_2$ 
in the Hamiltonian of a binary system was to compute (when considering, say,  the term linear in ${\mathbf S}_1$)
the spin precession angular velocity of  ${\mathbf S}_1$  in the gravitational field generated by the two masses $m_1, m_2$, and, eventually, the spin ${\mathbf S}_2$.
Indeed, this spin precession angular velocity (which can be obtained by writing that   ${\mathbf S}_1$ is parallely propagated along the
world line of $m_1$) is simply equal to the coefficient   ${\bfOmega}_1^{\rm SO}$ of  ${\mathbf S}_1$ in  $H_{\rm int}$.
On the other hand, it was recently remarked \cite{Dolan:2013roa,Bini:2014ica}  that, in the simple case of  a binary moving on circular orbits, the 
($z$-component of the) spin precession, $\Omega_1^{\rm SO}$, could be expressed in terms of the norm $|\nabla k|$ of the covariant derivative of the
helical Killing vector $k=\partial_t +\Omega \partial_\phi$  characteristic of circular motions, namely
\beq
\label{Omega1SO}
\Omega_1^{\rm SO}=\Omega - |\nabla k|\,,
\eeq
where $\Omega$ denotes the orbital frequency.
[The gauge-invariant quantity $|\nabla k|$
can be viewed as a first-derivative-level generalization of Detweiler's redshift invariant  \cite{Detweiler:2008ft}, which is expressible in terms of the norm $|k|$ of the
Killing vector $k$.] 

The gauge-invariant functional relation between $\Omega_1^{\rm SO}$, or equivalently   $|\nabla k|$, and the orbital frequency $\Omega$ has been recently studied
(both numerically and analytically)  in Refs. \cite{Dolan:2013roa,Bini:2014ica}. In particular, we have derived (as part of a sequence of analytical
gravitational self-force studies) in \cite{Bini:2014ica}
 the first-order self-force contribution (linear in the mass ratio $q=m_1/m_2 \ll 1$) to the \lq\lq geodetic" spin precession frequency $\Omega_1^{\rm SO}$
 to the eight-and-a-half post-Newtonian (PN) order, i.e. up to terms of order $y^{8.5}$ included, where 
 \beq
\label{3.6}
y=\left(\frac{G m_2\Omega}{c^3}\right)^{2/3}
\eeq
is a convenient dimensionless frequency parameter  of order $O(1/c^2)$. [We henceforth use, for simplicity, units where $G=c=1$.]
As in   \cite{Bini:2014ica} we restrict ourselves here to the case of a small spinning body $m_1, {\mathbf S}_1$, orbiting a large non-spinning body
$ m_2$, ${\mathbf S}_2 = 0$.

The aim of the present note is to report on an extension of our previous analytical computation  of spin precession to the 10.5PN level, i.e. up to terms of order $y^{10.5}$ included.
This extension was motivated by private communications from Dolan et al. \cite{Dolan_priv_com} who pointed out apparent discrepancies (starting at level $O(y^7)$) between some of their
 high-accuracy numerical results (see Table III in Ref. \cite{Dolan:2014pja}) and our published 8.5PN analytical results. These discrepancies led us to carefully re-examine our previous
computations, and to push them to higher PN orders. We so discovered that, though all our basic analytical building blocks were correct, their manipulation
by an algebraic software led to some instabilities (due to the length of the analytical expressions at high PN orders), that had led to a few errors
in our final results. More precisely, the rational term, among the seven  (transcendental) contributions to the coefficient  of $y^7$, was incorrectly obtained, 
and, in the coefficient of $y^8$ (which contains fifteen different contributions), both  
the rational term and the coefficient of $\pi^2$ were incorrectly obtained. Correspondingly, there were errors in the (rational) coefficients of $y^7$ and $y^8$ in the subtraction term $B(y)$. [See detailed results below.]  After having found these errors, corrected them, and
communicated the corrections to Dolan et al., the latter authors confirmed that our $O(y^8)$ corrected results were now in satisfactory agreement with their
high-accuracy numerical results.  [More recently,  Shah \cite{Shah_priv_com} independently pointed out to us the three discrepant coefficients mentioned above, which we had
already analytically derived, and which he and his collaborators had independently derived by using the numerical-analytical method of Ref. \cite{Shah:2013uya}.]

\section{Technical reminders}

Let us recall the notation and main technical results of  Ref. \cite{Bini:2014ica} that we shall need to express our new results.
We consider a two-body system  of masses $m_1$ and $m_2$, moving along circular orbits, in the limit $m_1\ll m_2$. Here we only endow the small mass $m_1$ with spin $S_1$, keeping the large mass $m_2$ non-spinning. 
This means that one is dealing with linear perturbations $h_{\mu\nu}(x^\lambda)$ of a Schwarzschild background of mass $m_2$ by a small mass $m_1$, moving on a circular orbit of radius $r_0$. 
As emphasized by  Detweiler \cite{Detweiler:2008ft},  the perturbed metric admits the helical Killing vector $k=\partial_t +\Omega \partial_\phi$, i.e.,  the metric perturbation depends only on $\bar \phi=\phi-\Omega t$, $r$ and $\theta$, $h_{\mu\nu}(\bar \phi , r, \theta)$. 

The four-velocity of $m_1$, normalized with respect to the metric $g^R_{\mu\nu}(x^\lambda)=g^{(0)}_{\mu\nu}+q\, h_{\mu\nu}^{\rm R}+O(q^2)$, (here $q\equiv m_1/m_2\ll 1$ and  the superscript $R$ indicates the {\it regular} part \cite{Detweiler:2002mi} of  $h_{\mu\nu}(x^\lambda)$ around the world line of $m_1$), can be written as
\beq
\label{3.1}
U_1^\mu=\frac{k^\mu}{|k|}\equiv \Gamma k^\mu\,,\qquad \Gamma \equiv \frac{1}{|k|}\,,
\eeq
where (to linear order in $q$)
\begin{eqnarray}
\label{3.2}
|k| &=& \sqrt{[-g_{\mu\nu}^{\rm R} k^\mu k^\nu]_1}=\sqrt{1-\frac{2m_2}{r_0}-\Omega^2 r_0^2-qh_{kk}}\nonumber\\
&=& \sqrt{1-\frac{2m_2}{r_0}-\Omega^2 r_0^2}\left(1-\frac{1}{2}q \frac{h_{kk}}{1-\frac{2m_2}{r_0}-\Omega^2 r_0^2} \right)\nonumber\\
\end{eqnarray}
with $h_{kk}=[h_{\mu\nu}^{\rm R}(x)k^\mu k^\nu]_1$. 
Writing that $m_1$ moves along an equatorial  circular geodesic yields the  conditions $\partial_\mu g^R_{kk}=0$, which lead to  \cite{Detweiler:2008ft}
\begin{eqnarray}
\label{3.4}
\Omega &=& \sqrt{\frac{m_2}{r_0^3}}\left(1-q \frac{r_0^2}{4m_2}[\partial_r h_{kk}^{\rm R}]_1  \right)\,,\\
\label{3.5}
[\partial_{\bar \phi} h_{kk}^{\rm R}]_1&=&0\,.
\end{eqnarray}
Eq. (\ref{3.4})  allows one to trade the gauge-dependent radius $r_0$ for the gauge-invariant dimensionless frequency parameter $y$, Eq. (\ref{3.6}), using 
\begin{eqnarray}
\label{3.7}
r_0 &=& \frac{m_2}{y}-q \frac{m_2^2}{6y^3}[\partial_r h_{kk}^{\rm R}]_1\,,\nonumber\\
\label{3.8}
\frac{m_2}{r_0}&=&y \left(1+q \frac{m_2}{6y^2}[\partial_r h_{kk}^{\rm R}]_1 \right)\,.
\end{eqnarray}

The geodetic spin-orbit precession frequency along the world line of $m_1$  
has, as only nonvanishing component, $\Omega^{\rm SO}_1\equiv \Omega_z^{\rm SO}$ given by Eq. (\ref{Omega1SO}) above.
In this equation, 
the norm $|\nabla k|$ of the covariant derivative of the
helical Killing vector $k=\partial_t +\Omega \partial_\phi$ is defined as
\beq
|\nabla k|^2 =\frac12  (\nabla_\mu k_\nu) (\nabla^\mu k^\nu)\,,   
\eeq
where all tensorial operations are done with the metric $g_{\mu\nu}^{\rm R}(x)$.
The  explicit expression of $|\nabla k |$
can be written as
\beq
\label{3.9}
|\nabla k|=|\nabla k|^{(0)}\, (1+q\, \delta(y)+O(q^2))\,,
\eeq
where
\beq
\label{3.11}
|\nabla k|^{(0)}=\Omega \sqrt{1-3y}\,,
\eeq
is the well-known result for gyroscopic precession (with respect to a rotating, polar-coordinate frame) in a Schwarzschild background
\cite{Straumann}, and where 
\begin{eqnarray}
\label{3.10}
\delta (y) &=&  
 -\frac12 (1-2y)h_{rr}-\frac{y^2(1-y)}{2m_2^2 (1-2y)}h_{\phi\phi}\nonumber\\
&& -\frac{y^{3/2}}{m_2(1-2y)}h_{t\phi}-\frac{y}{2(1-2y)(1-3y)}h_{kk}\nonumber\\
&& -\frac{1}{2\sqrt{y}}(\partial_{\phi} h_{rk}-\partial_r h_{\phi k})  \,.
\end{eqnarray}
In Eq. (\ref{3.10}) all quantities are to be regularized and evaluated for $\theta=\pi/2$.

The quantity $\delta(y)$, which measures the fractional first order self-force (1SF)  correction to $|\nabla k|$, is equivalent to the quantity $\delta \psi(y)$
which measures the 
1SF contribution to the dimensionless ratio \cite{Dolan:2013roa}
\beq
\label{5.1}
\psi(y)\equiv \frac{\Omega_1^{\rm SO}}{\Omega}=1-\frac{|\nabla k|}{\Omega}=1-\sqrt{1-3y}[1+q\, \delta(y)+O(q^2)]\,.
\eeq 
Explicitly, we have
\beq
\label{5.3}
\delta \psi (y)=-\sqrt{1-3y}\, \delta(y)\,.
\eeq 
 
Following the methodology explained in Refs. \cite{Bini:2013zaa, Bini:2013rfa,Bini:2014nfa,Bini:2014zxa,Bini:2015bla},
and extending the results of Ref. \cite{Bini:2014ica} to higher post-Newtonian orders (by using radiative solutions, $X_{\rm (in)}$, $X_{\rm (up)}$,  up to $l=7$), we have computed $\delta(y)$ up to order $y^{10.5}$.

\section{New higher post-Newtonian terms in $\delta(y)$ and $\delta \psi (y)$} 

Before listing the 
 complete expressions of $\delta(y)$ and $\delta \psi (y)$ to order  $y^{10.5}$ let us indicate that our previous $O(y^{8.5})-$accurate results 
missed one term at  level $y^7$ and two terms
at  level $y^8$, while the $y^{7.5}$ and $y^{8.5}$ terms were complete.

More precisely, the correct $O(y^{8.5})-$accurate   expression of $\delta(y)$ is  obtained by adding $\Delta c^\delta_7\, y^7+\Delta c^\delta_8\, y^8$ to Eq. (4.33) in  \cite{Bini:2014ica},
where
\begin{eqnarray}
\label{diff_delta}
\Delta c^\delta_7&=&-\frac{1485630311863}{45831035250}\,,\nonumber\\
\Delta c^\delta_8&=&\frac{8}{8505}\pi^2-\frac{25377697082469367}{262980313505910}\,.
\end{eqnarray}
Equivalently, the correct $O(y^{8.5})-$accurate   expression of $\delta\psi (y)$ is  obtained by adding $\Delta c^{\delta\psi}_7 y^7+\Delta c^{\delta\psi}_8 y^8$ to Eq. (5.4) in  \cite{Bini:2014ica}, where $\Delta c^{\delta\psi}_7=-\Delta c^\delta_7$ and $\Delta c^{\delta\psi}_8=-\Delta c^\delta_8+\frac32\,\Delta c^\delta_7$, i.e.,
\begin{eqnarray}
\label{diff_delta_psi}
\Delta c^{\delta\psi}_7 &=&\frac{1485630311863}{45831035250}\nonumber\\
\Delta c^{\delta\psi}_8 &=& -\frac{8}{8505}\pi^2+\frac{629539392522290711}{13149015675295500}\,.
\end{eqnarray}

\begin{widetext}
The full $O(y^{10.5})-$accurate expressions of $\delta(y)$ and $\delta \psi (y)$  read
\begin{eqnarray}
\delta(y)&=&-y^2+\frac32 y^3 +\frac{69}{8}y^4+\left(\frac{53321}{240}+\frac{496}{15}\ln(2)+16\gamma+8\ln(y)-\frac{20471}{1024}\pi^2 \right) y^5\nonumber\\
&+& \left(\frac{15462423}{4480}+\frac{172}{5}\gamma+\frac{1436}{105}\ln(2)-\frac{357521}{1024}\pi^2+\frac{86}{5}\ln(y)+\frac{729}{14}\ln(3)\right) y^6\nonumber\\
&+& \frac{26536}{1575} y^{13/2}\pi\nonumber\\
&+& \left(\frac{16156122817}{1209600}-\frac{30832}{105}\gamma-\frac{3344}{21}\ln(2)-\frac{512537515}{393216}\pi^2-\frac{15416}{105}\ln(y)-\frac{40581}{140}\ln(3)+\frac{1407987}{524288}\pi^4\right) y^7\nonumber\\
&+&\frac{670667}{22050} y^{15/2}\pi\nonumber\\
&+&\left(-\frac{41432062371919}{2540160000}+\frac{96697099}{141750}\gamma-\frac{58208}{105}\ln(2)^2-\frac{1291394011}{3638250}\ln(2)-\frac{1007542476707}{353894400}\pi^2\right. \nonumber\\
&+& \frac{9765625}{28512}\ln(5)+\frac{96697099}{283500}\ln(y)+\frac{2364633}{12320}\ln(3)-\frac{856}{25}\ln(y)^2+\frac{162286431837}{335544320}\pi^4-\frac{869696}{1575}\ln(2)\gamma\nonumber\\
&+&\left.\frac{1344}{5}\zeta(3)-\frac{3424}{25}\gamma\ln(y)-\frac{3424}{25}\gamma^2 -\frac{434848}{1575}\ln(2)\ln(y)\right) y^8\nonumber\\
&-&\frac{3872542979}{13097700} y^{17/2}\pi\nonumber\\
&+&\left(-\frac{4084955265168837911}{1173553920000}+\frac{118580138377}{14553000}\gamma+\frac{45728}{1225}\ln(2)^2+\frac{58794404629417}{3972969000}\ln(2)\right. \nonumber\\
&+& \frac{7776}{5}\zeta(3)-\frac{100335874551071}{26424115200}\pi^2 -\frac{20486328125}{5189184}\ln(5)+\frac{118580138377}{29106000}\ln(y)\nonumber\\
&-& \frac{32288}{75}\gamma^2+\frac{143985009429}{15695680}\ln(3)-8072/75\ln(y)^2+\frac{773697968441461}{21474836480}\pi^4\nonumber\\
&-& \frac{28431}{49}\ln(y)\ln(3)-\frac{56862}{49}\gamma\ln(3)-\frac{56862}{49}\ln(2)\ln(3)\nonumber\\
&-& \left.\frac{1210816}{2205}\ln(2)\gamma-\frac{32288}{75}\gamma\ln(y)-\frac{605408}{2205}\ln(2)\ln(y)-\frac{28431}{49}\ln(3)^2
\right) y^9\nonumber\\
&+& \left(\frac{460314955849127}{524431908000}\pi-\frac{6228256}{11025}\ln(2)\pi-\frac{23264368}{165375}\pi\ln(y)+\frac{434848}{4725}\pi^3-\frac{46528736}{165375}\pi\gamma\right) y^{19/2}\nonumber\\
&+& \left(-\frac{5405869945189728461825461}{169160756244480000}-\frac{164976460027543}{15891876000}\gamma-\frac{4653978748}{467775}\ln(2)^2\right. \nonumber\\
&-& \frac{164366211989143}{31783752000}\ln(y)+\frac{144656188561370737}{4719887172000}\ln(2)+\frac{1337603}{2205}\ln(y)^2-\frac{489993464291995}{532710162432}\pi^2\nonumber\\
&-&\frac{80728}{21}\zeta(3)+\frac{5350412}{2205}\gamma\ln(y)-\frac{1169541476}{3274425}\ln(2)\ln(y)-\frac{2339082952}{3274425}\ln(2)\gamma+\frac{5350412}{2205}\gamma^2\nonumber\\
&+& \frac{86209353}{26950}\ln(3)^2-\frac{167271372501741}{3453049600}\ln(3)+\frac{161107421875}{10378368}\ln(5)+\frac{906012273831305533}{2748779069440}\pi^4\nonumber\\
&-& \left. \frac{21138410295}{134217728}\pi^6+\frac{86209353}{26950}\ln(y)\ln(3)+\frac{86209353}{13475}\gamma\ln(3)+\frac{86209353}{13475}\ln(2)\ln(3)+\frac{678223072849}{370656000}\ln(7)\right) y^{10}\nonumber\\
&+& \left(\frac{1242850565271443431}{159077678760000}+\frac{3164198}{6615}\pi^2-\frac{60364562}{77175} \gamma-\frac{26484566}{55125}\ln(2) -\frac{369603}{343} \ln(3)-\frac{30182281}{77175} \ln(y)\right) \pi y^{21/2} \nonumber\\
&+& O_{\ln}(y^{11})\,.
\end{eqnarray}

\begin{eqnarray}
\delta \psi &=&  y^2-3y^3-\frac{15}{2}y^4+\left(-\frac{6277}{30}-\frac{496}{15}\ln(2)-16\gamma-8\ln(y)+\frac{20471}{1024}\pi^2\right) y^5\nonumber\\
&&+\left(-\frac{87055}{28}+\frac{3772}{105}\ln(2)-\frac{52}{5}\gamma-\frac{26}{5}\ln(y)+\frac{653629}{2048}\pi^2-\frac{729}{14}\ln(3)\right) y^6\nonumber\\
&&-\frac{26536}{1575} y^{13/2}\pi\nonumber\\
&&+\left(-\frac{149628163}{18900}+\frac{7628}{21}\gamma+\frac{3814}{21}\ln(y)+\frac{4556}{21}\ln(2)+\frac{12879}{35}\ln(3)-\frac{1407987}{524288}\pi^4+\frac{297761947}{393216}\pi^2\right) y^7\nonumber\\
&&-\frac{113411}{22050} y^{15/2}\pi\nonumber\\
&&+\left(-\frac{74909462}{70875}\gamma+\frac{58208}{105}\ln(2)^2+\frac{340681718}{1819125}\ln(2)+\frac{164673979457}{353894400}\pi^2\right. \nonumber\\
&& -\frac{160934764317}{335544320}\pi^4-\frac{1344}{5}\zeta(3)+\frac{869696}{1575}\ln(2)\gamma+\frac{3424}{25}\gamma^2-\frac{199989}{352}\ln(3)\nonumber\\
&& -\frac{9765625}{28512}\ln(5)+\frac{856}{25}\ln(y)^2+\frac{434848}{1575}\ln(2)\ln(y)-\frac{37454731}{70875}\ln(y)\nonumber\\
&&\left. +\frac{403109158099}{9922500}+\frac{3424}{25}\gamma\ln(y)\right)y^8\nonumber\\
&&
+\frac{1179591206}{3274425}y^{17/2}\pi\nonumber\\
&& +\left(-\frac{4454779894}{606375}\gamma-\frac{1064368}{1225}\ln(2)^2-\frac{138895624334}{9029475}\ln(2)-\frac{22832200546571}{8808038400}\pi^2\right. \nonumber\\
&& -\frac{758053590944149}{21474836480}\pi^4-1152\zeta(3)-\frac{3077728}{11025}\ln(2)\gamma+\frac{3376}{15}\gamma^2+\frac{28431}{49}\ln(3)^2\nonumber\\
&& -\frac{71602663581}{7847840}\ln(3)+\frac{11576171875}{2594592}\ln(5)+\frac{844}{15}\ln(y)^2+\frac{28431}{49}\ln(y)\ln(3)\nonumber\\
&& -\frac{1538864}{11025}\ln(2)\ln(y)-\frac{2227389947}{606375}\ln(y)+\frac{56862}{49}\gamma\ln(3)+\frac{3985926908910281}{1146048750}\nonumber\\
&& \left. +\frac{56862}{49}\ln(2)\ln(3)+\frac{3376}{15}\gamma\ln(y)\right) y^9\nonumber\\
&&+\left(-\frac{660044682996077}{524431908000}\pi-\frac{434848}{4725}\pi^3+\frac{46528736}{165375}\pi\gamma+\frac{6228256}{11025}\ln(2)\pi\right. \nonumber\\
&& \left. +\frac{23264368}{165375}\pi\ln(y)\right) y^{19/2}\nonumber\\
&&+\left(\frac{11467229058074}{496621125}\gamma+\frac{21138410295}{134217728}\pi^6+\frac{30719079112}{3274425}\ln(2)^2-\frac{1306135539288758}{147496474125}\ln(2)\right. \nonumber\\
&&-\frac{152033994681460553}{13317754060800}\pi^2-\frac{755954175166870909}{2748779069440}\pi^4+\frac{680336}{105}\zeta(3)-\frac{478423984}{654885}\ln(2)\gamma\nonumber\\
&& -\frac{35570296}{11025}\gamma^2-\frac{54832464}{13475}\ln(3)^2+\frac{214411899501351}{3453049600}\ln(3)-\frac{437134765625}{20756736}\ln(5)\nonumber\\
&& -\frac{8892574}{11025}\ln(y)^2-\frac{54832464}{13475}\ln(y)\ln(3)-\frac{239211992}{654885}\ln(2)\ln(y)+\frac{5724079403437}{496621125}\ln(y)\nonumber\\
&& -\frac{109664928}{13475}\gamma\ln(3)+\frac{552424223705497767347}{20649506377500}-\frac{109664928}{13475}\ln(2)\ln(3)-\frac{35570296}{11025}\gamma\ln(y)\nonumber\\
&& \left. -\frac{678223072849}{370656000}\ln(7)\right) y^{10}\nonumber\\
&&+\left(-\frac{178279193702345741}{26512946460000}\pi-\frac{11255086}{33075}\pi^3+\frac{46324078}{128625}\pi\gamma-\frac{2889622}{7875}\ln(2)\pi\right. \nonumber\\
&& \left. +\frac{369603}{343}\pi\ln(3)+\frac{23162039}{128625}\pi\ln(y)\right) y^{21/2}+ O_{\ln}(y^{11})\,.
\end{eqnarray}

\end{widetext}
Note that, in the above expressions, we used the computer-algebra-related notation $\ln(a)^n$ to denote $\ln^n (a)$ and $O_{\ln}(y^{11})$ to denote a term of order 
$y^{11}\ln^n y$ for some $n$. The corresponding $O_{\ln{}}(u^{11})-$accurate expansion of the effective gyrogravitomagnetic ration $g_{S_*}^{\rm 1SF}(u)$ is given in the Appendix.

\section{Analytic expression of the subtraction term}

Prompted by  Dolan et al. \cite{Dolan_priv_com}, who pointed out discrepancies at order $y^7$ and $y^8$ between our 
 Eq. (4.30) in \cite{Bini:2014ica} and their (unpublished) corresponding expression for the subtraction term $B(y)$, we have found a way to
derive an exact analytic expression for $B(y)$ within our formalism, which is based on  Regge-Wheeler-Zerilli-type {\it tensorial} multipolar expansions.
As we shall now explain, our derivation is a novel approach grounded on a Wentzel-Kramers-Brillouin (WKB)  analysis 
of the homogeneous radial (Regge-Wheeler, RW)  equation satisfied by the fundamental building blocks, $X_{\rm in}$ and $X_{\rm up}$, of our 
formalism.
This WKB approach (which we explain in detail below) is quite different from the approach traditionally used in gravitational self-force theory,
which is based on local, Hadamard-type expansions of the metric $h_{\mu\nu}$, in Lorenz-gauge, near the world line of $m_1$ (see e.g., \cite{Detweiler:2002mi,Barack:2001gx,Detweiler:2002gi}). In addition, our approach
defines the subtraction terms by considering the limit $l\to \infty$ where $l$ denotes the degree in a {\it tensorial} multipolar expansion, while
the usual self-force calculations define subtraction terms by considering a limit $l_s\to \infty$, where $l_s$ denotes the order in a  {\it scalar} multipolar expansion.
One can  show that, for the quantities we shall consider, the two different limiting procedures should give the same subtraction terms at leading order. [However, at higher orders in local singularity expansions, the extension ambiguities of such expansions do not imply anymore their equivalence.]

Let us start by recalling the form of the WKB approximation of the solutions of a one-dimensional Schr\"odinger equation, say  
\beq
\label{Schro}
\frac{d^2}{dx^2}\Psi=\frac{Q(x)}{\hbar^2}\Psi(x)\,.
\eeq
The WKB solutions of Eq. (\ref{Schro}) are written in the form
\beq
\label{WKBexp}
\Psi(x)=e^{\frac{S_0}{\hbar}+S_1 +O(\hbar)}\,.
\eeq
As indicated here, it will be sufficient for our purpose to keep only the leading and next-to-leading terms in the WKB expansion.
At this order of approximation, the  two independent solutions of Eq. (\ref{Schro}) read
\beq
\label{WKB_sol}
\Psi_\pm(x)=C_\pm \frac{e^{\pm \int p(x)dx}}{\sqrt{p(x)}}\,,\qquad  p(x)=\sqrt{Q(x)}\,,
\eeq
corresponding to
\beq
\frac{S_0}{\hbar}=\pm \int p(x) dx\,,\qquad S_1=-\frac12 \ln p(x)\,.
\eeq	
The choice $C_\pm=1/\sqrt{2}$ would imply that the Wronskian of these solutions is 1:
\beq
W=\Psi_- \Psi_+'-\Psi_+ \Psi_-'=1\,.
\eeq
Note that we will use the WKB approximation in the classically forbidden domain, where $Q(x)$ is positive so that
the solutions $\Psi_\pm$ are exponentially growing or decaying.

We  first apply this approximation to the (homogeneous) radial RW equation
\beq
\label{RW_eq}
\frac{d^2}{dr_\ast{}^2}X=\left[ f(r)\left(\frac{l(l+1)}{r^2}-\frac{6M\eta^2}{r^3}\right)-\eta^2 m^2 \Omega^2 \right]X\,.
\eeq
Here, $\eta\equiv 1/c$, $m$ is the spherical harmonics order, $\Omega$ denotes the orbital frequency, and
\beq
dr_\ast=\frac{dr }{f(r)}\,,\quad f(r)=1-\frac{2M\eta^2}{r}\,.
\eeq
The spatial variable (denoted $x$ in Eq.(\ref{Schro})) in this one-dimensional Schr\"odinger equation is $r_\ast$, while we shall take as small expansion parameter $\hbar$ the quantity
\beq
\hbar\equiv \frac{1}{ L }\,,
\eeq
where we introduced the convenient notation
\beq
L=l+\frac12\,.
\eeq
Note indeed that the coefficient $l(l+1)$ in the centrifugal potential can be written as
\beq
l(l+1)=L^2-\frac14\,,
\eeq
and is of order $\sim \frac{1}{\hbar^2}$. 

In order to capture the near-world-line singularity expansion within our tensorial multipolar expansion, we need to consider a limit where
both $l\sim L$ and $m$ tend  to infinity with the  ratio $w\equiv m/L$ being kept fixed.
In this limit the two dominant terms (of order $1/\hbar^2$) in Eq. (\ref{Schro})  are 
\begin{eqnarray}
\frac{Q_\ast}{\hbar^2}&=& l(l+1) \left[\frac{1}{r^2} f(r)  -\eta^2 \frac{m^2}{l(l+1)} \Omega^2  \right]+ O(L^0)  \nonumber\\
&=& L^2\left[\frac{1}{r^2} f(r)  -\eta^2 \frac{m^2}{L^2} \Omega^2  \right]+ O(L^0)\nonumber\\
&=& L^2\left[\frac{1}{r^2}  f(r) -\eta^2 w^2 \Omega^2  \right]+ O(L^0)\,.
\end{eqnarray}
Correspondingly to the accuracy used in Eq. (\ref{WKBexp}), we can neglect the terms of order $O(L^0)$ in the above equation, which notably means
neglecting the term $6M/r^3$ in Eq. (\ref{RW_eq}). At this stage no expansion is performed in the PN-parameter $\eta=1/c$.

Introducing the notation
\beq
\Delta(r)=  1-\frac{2M\eta^2}{r} -\eta^2 w^2 \Omega^2 r^2\,,
\eeq
we have
\beq
Q_\ast =p_\ast^2=L^2 \frac{\Delta(r)}{r^2}\,,\qquad p_\ast=  L \frac{\sqrt{\Delta(r)}}{r}\,,
\eeq
so that the building blocks of the WKB solution \eqref{WKB_sol} read
\beq
\frac{S_0}{\hbar}=\pm \int p_\ast dr_\ast=\pm \int p_\ast \frac{dr}{f}=\pm L\int \frac{\sqrt{\Delta(r)}}{f(r)}\frac{dr}{r}\,,
\eeq
and
\beq
S_1=-\frac12 \ln p_\ast\,.
\eeq
More explicitly
\beq
\sqrt{p_\ast}=\sqrt{L}\frac{\Delta^{1/4}}{\sqrt{r}}
\eeq
so that
\beq
\frac{C_\pm }{\sqrt{p_\ast}}= \tilde C_\pm  \frac{\sqrt{r}}{\Delta^{1/4}}    
\eeq
where we have re-absorbed the factor $\sqrt{L}$ in the constant $C_\pm$ [$\tilde C_\pm=C_\pm/\sqrt{L}$].

The final result of this WKB analysis is that two independent solutions of the RW equation  \eqref{RW_eq} are
\beq
X_\pm =\tilde C_\pm  \frac{\sqrt{r}}{\Delta^{1/4}} e^{\pm L \int \frac{\sqrt{\Delta}}{f}\frac{dr}{r}}\,.
\eeq
We  checked  
that the PN expanded solutions of the RW equation that we constructed in our formalism \cite{Bini:2013zaa, Bini:2013rfa, Bini:2014nfa, Bini:2014zxa, Bini:2015bla} agree with those  WKB solutions, with the following correspondence
\begin{eqnarray}
\label{R_in_up_WKB}
X_{\rm (in)}&\approx &\frac{\sqrt{r}}{\Delta^{1/4}}e^{L\int \frac{\sqrt{\Delta}}{rf}dr} \,,\nonumber\\
X_{\rm (up)}&\approx &\frac{\sqrt{r}}{\Delta^{1/4}}e^{-L\int \frac{\sqrt{\Delta}}{rf}dr} \,.
\end{eqnarray}
Note for instance that, when expanding in powers of $\eta$ the right-hand-side (rhs) of $X_{\rm (in)}$,  as given in Eq. \eqref{R_in_up_WKB}, 
its leading order is $\sqrt{r}e^{L\ln r}=r^{l+1}$ in agreement with the normalization of our PN solution which was chosen as
\beq
X_{\rm (in)}^{l\omega}(r)=r^{l+1}\left(1+A^{l\omega}(r)  \right)\,,
\eeq
with $A^{l\omega}(r)=O(\eta^2)$.

Inserting the above WKB solutions for $X_{\rm (in)}$ and $X_{\rm (up)}$ in the analytical expressions for $\delta_{lm}^{\pm \rm (odd/even)}$ given in  Ref. \cite{Bini:2014ica} [see Eqs. (4.10) and (4.12) together with Eqs. (4.11), (4.13) and (4.23) there] yields 
expressions  for $\delta_l^\pm (y)=\delta_l^{\pm, \rm even} (y)+\delta_l^{\pm \rm odd} (y) $ of the form of Eq. (4.28) there, i.e.,
\beq
\label{delta_pm}
\delta_l^\pm (y)\equiv \sum_m \delta_{lm}^\pm(y)=\pm L \, A(y)+ B(y)+O\left(\frac{1}{L^2} \right)\,.
\eeq
At this stage the subtraction  term $B(y)$ is given by a sum over $m$ of the form
\beq
\sum_m f\left(\frac{m}{L} \right)|Y_{lm}|^2+g\left(\frac{m}{L} \right)\left|\frac{d Y_{lm}}{d\theta}\right|^2\,,
\eeq 
where $Y_{lm}(\theta, \phi)$ and its $\theta-$derivative are both evaluated at $\theta=\pi/2$ (and $\phi=0$).
Such a  sum can be asymptotically evaluated, in the limit $L\to \infty$ with $m/L$ fixed, in terms of an integral, between $-1$ and $1$, over the variable $w=m/L$.
In order to do so one needs asymptotic estimates for $|Y_{lm}|^2$ and $\left|\frac{d Y_{lm}}{d\theta}\right|^2$ as functions of $w$ in the large $L$ limit.
Such asymptotic estimates can be derived by a WKB analysis of the $\theta$ differential equation satisfied by $\Theta_{lm}(\theta)$ (defined by factoring $Y_{lm}(\theta, \phi)=\Theta_{lm}(\theta) e^{im\phi}$). Indeed, $\Theta_{lm}(\theta)$ satisfies a one-dimensional Schr\"odinger equation of the type \eqref{Schro}, when using the variable $\lambda=\int_{\pi/2} d\theta/\sin \theta$, namely
\beq
\frac{d^2}{d\lambda^2}\Theta_{lm}=-P^2(\lambda) \Theta_{lm}\,,
\eeq
with
\beq
P^2(\lambda)= l(l+1) \sin^2 \theta (\lambda) -m^2\,.
\eeq
This leads to WKB solutions of the type
\beq
\Theta_{lm}(\lambda)=C_+ \frac{e^{i\int P(\lambda) d\lambda}}{\sqrt{P(\lambda)}}+C_- \frac{e^{-i\int P(\lambda) d\lambda}}{\sqrt{P(\lambda)}}\,,
\eeq 
for appropriate choices of the constants $C_\pm$ determined by regularity conditions at $\lambda=- \infty$ (corresponding to $\theta=0$) and 
$\lambda=+ \infty$ (corresponding to $\theta=\pi$). When evaluating $\Theta_{lm}(\lambda)$ and $\frac{d}{d\lambda}\Theta_{lm}(\lambda)=\sin \theta \frac{d}{d\theta}\Theta_{lm}$ at $\theta=\pi/2$ (i.e., $\lambda=0$) one finds the following WKB estimates
\beq
\label{Ysquare}
|Y_{lm}(\frac{\pi}{2},0)|^2\approx \frac{1}{\pi^2}\frac{1}{\sqrt{1-w^2}}\delta^{\rm even}_{l-m}\,,
\eeq
and
\beq
\label{DYsquare}
|\partial_\theta Y_{lm}(\frac{\pi}{2},0)|^2\approx \frac{L^2}{\pi^2} \sqrt{1-w^2} \delta^{\rm odd}_{l-m}\,.
\eeq
Here $\delta^{\rm even}_{l-m}$ ($\delta^{\rm odd}_{l-m}$) is equal to 1 when $l-m$ is even (odd) and to 0 otherwise.
The above estimates are not  a priori uniformly valid in the full range $-l\le m \le l$ because our WKB  analysis requires $L^2-m^2\gg 1$.
However, they can correctly evaluate the asymptotic values of the integrals that we shall be interested in below (which have only a relatively small contribution from the neighborhoods of the boundary points $w=\pm 1$).
We have checked the estimates \eqref{Ysquare} and \eqref{DYsquare} by using the explicit expressions of  $Y_{lm}(\frac{\pi}{2},0)$ and $\partial_\theta Y_{lm}(\frac{\pi}{2},0)$ given in Eqs. (32) and (33) of \cite{Shah:2010bi}\,. 
A consequence of Eqs. \eqref{Ysquare} and \eqref{DYsquare} is that
\begin{eqnarray}
\sum_m \frac{4\pi}{2l+1} |Y_{lm}(\frac{\pi}{2},0)|^2 f\left( \frac{m}{L}\right)&\approx & \frac{1}{\pi}\int_{-1}^1 \frac{dw}{\sqrt{1-w^2}} f(w)\,, \nonumber\\
\sum_m \frac{4\pi}{2l+1} \frac{|\partial_\theta Y_{lm}(\frac{\pi}{2},0)|^2}{L^2} g\left( \frac{m}{L}\right)&\approx & \frac{1}{\pi}\int_{-1}^1 dw \sqrt{1-w^2} g(w)\nonumber\\ \,.
\end{eqnarray}
As an example of the application of these asymptotic estimates we have computed the analytic expression of the $L\to \infty$ limit of the first-order self-force redshift quantity $h_{kk}$. Starting from Eqs. (29) and (30) of Ref. \cite{Bini:2013zaa} one finds that $B_{h_{kk}}\equiv \lim_{l\to \infty} (h_{kk,lm}^{\rm (even)}+h_{kk,lm}^{\rm (odd)})$ is given by
\begin{eqnarray}
\label{hkk_limit}
B_{h_{kk}}
&=& \frac{2y (1-3y)^{3/2}}{\sqrt{1-2y}}\frac{1}{\pi}\int_{-1}^1 \frac{dw}{\sqrt{(1-w^2)(1-k^2 w^2)}}\nonumber\\
&=& \frac{2y (1-3y)^{3/2}}{\sqrt{1-2y}} \frac{2}{\pi}{\rm EllipticK}(k)\,,
\end{eqnarray}
where
\beq
k^2=\frac{y}{1-2y}\,,
\eeq
and where ${\rm EllipticK}(k)$ denotes the complete elliptic integral of the first kind (with $w\equiv \sin \alpha$):
\beq
{\rm EllipticK}(k)=\int_0^{\pi/2} \frac{d\alpha}{\sqrt{1-k^2 \sin^2\alpha}}\,.
\eeq
This result agrees with the subtraction term obtained by the usual self-force Hadamard-type analysis \cite{Detweiler:2002gi,Detweiler:2008ft}, i.e., the term denoted 
$\tilde D_0=(1-3y)D_0$ in \cite{Bini:2013zaa,Bini:2013rfa}. [Note that there is a misprint in the last term of Eq. (56) in \cite{Bini:2013rfa}; the coefficient of $u^7$ should read $+4409649/524288$].

When applying the above WKB asymptotic estimates (for both the radial functions $X_{\pm}(r)$ and the angular functions $Y_{lm}$ and $\partial_\theta Y_{lm}$) to the $l\to \infty$ limit of  the quantity $\delta^\pm_l(y)$, \eqref{delta_pm}, we obtain the following analytic expression for the $O(L^0)$ subtraction term $B(y)$
\begin{widetext}
\beq
B_{\rm WKB}(y)=\frac{1}{\pi}\sqrt{\frac{1-3y}{1-2y}}\left[(4-9y){\rm EllipticE}(k) -2(2-5y){\rm EllipticK}(k) \right]\,.
\eeq
\end{widetext}
Here ${\rm EllipticE}(k)$ denotes the complete elliptic integral of the second kind 
\beq
{\rm EllipticE}(k)=\int_0^{\pi/2}d\alpha\, \sqrt{1-k^2 \sin^2\alpha}  \,.
\eeq

The expansion in powers of $y$ of $B_{\rm WKB}$ reads, up to the $11$ PN level 
\begin{eqnarray}
\label{B_exp}
B_{\rm WKB}(y)&=& -\frac12  y+\frac14  y^2+\frac{63}{128} y^3+\frac{995}{1024} y^4+\frac{63223}{32768} y^5\nonumber\\
&& +\frac{126849}{32768} y^6+\frac{16567767}{2097152} y^7+\frac{555080733}{33554432} y^8\nonumber\\
&& +\frac{77104836855}{2147483648} y^9+\frac{350273500199}{4294967296} y^{10}\nonumber\\
&& +\frac{26812467118879}{137438953472} y^{11}+O(y^{12})\,.
\end{eqnarray}
In our previous work \cite{Bini:2014ica}  the subtraction term $B$ was not derived independently of our computation of $\delta_{lm}^\pm$ but was obtained 
from the large $l$ limit of the PN expanded version of $\delta_{lm}^\pm$. The algebraic-manipulation errors mentioned above induced corresponding errors in our previous evaluation of the PN expansion of $B$ (as pointed out to us by Dolan et al. \cite{Dolan_priv_com}). More explicitly, the coefficients of 
$y^7$ and $y^8$ in Eq. (4.30) in  \cite{Bini:2014ica} were in error,  and Eq. \eqref{B_exp} gives instead their correct values.

\section{Concluding remarks}
The analytic computation of the post-Newtonian expansion of the first-order self-force contribution to  spin precession has been raised here to the ten and ten-and-half
post-Newtonian level.

Our analysis has also corrected two terms (at the PN levels 7 and 8) among our previous 8.5 PN-accurate calculation of spin-orbit effects \cite{Bini:2014ica}.
More precisely, we have shown that  Eq. (4.33) in  \cite{Bini:2014ica} needs to be augmented by the two terms in Eq. \eqref{diff_delta}.
Equivalently,  Eq. (5.4) in  \cite{Bini:2014ica} needs to be augmented by the two terms in Eq. \eqref{diff_delta_psi}.
These missing terms were caused by algebraic errors in the manipulation of large analytic expressions. Note that these errors affected only a few terms among many contributions (essentially only rational terms). The missing contributions to the 
coefficients of $y^7$ and $y^8$ in $\delta\psi(y)$ are numerically equal to $\Delta c_7^\psi =32.41537757$ and $\Delta c_8^\psi =47.86801827$. These values are rather small compared to the corresponding typical values of the general PN coefficient $c_n^\psi\sim -0.12 \times 3^n$ (linked to the pole singularity of $\delta\psi(y)$ at $y=\frac13$, see Eq. (5.14) in  \cite{Bini:2014ica}).
The fractional modifications brought to the coefficients $g_6^c$ and $g_7^c$ in Eqs. (6.36), (6.37) and (6.38) are correspondingly small, $\delta g_6/g_6\simeq-0.02487821950$ and $\delta g_7/g_7\simeq 0.0001739775786$.
As a consequence, correcting these terms does not affect any of the significant conclusions we reached in \cite{Bini:2014ica} which were mainly aimed at describing strong field effects. In particular, our fits Eqs. (5.11) and  (6.39) did not make any use of the $y^7$ and $y^8$ coefficients but only relied on 3PN information and on the strong field  numerical data of \cite{Dolan:2013roa}.

Finally, we have introduced here a new method for analytically computing the subtraction terms of self-force quantities. Instead of the traditionally used Hadamard-like near-world-line singularity expansions, our new method is based on a WKB analysis of both the radial and angular equations satisfied by the metric, when considering them in the limit $l\to \infty$ with the ratio $m/l$ fixed. We have shown on two examples ($h_{kk}$ and the spin precession) that our method leads rather simply to closed form expressions for the subtraction terms.

\appendix
\section{Higher PN terms in $g_{S_*}^{\rm 1SF}$}

Combining the $O(y^{10.5})$-accurate computation of $\delta(y)$ above with our recent $O(y^{10.5})$-accurate computation of   the main effective one-body radial potential $a(u)$ \cite{Bini:2015bla}, we can raise the PN expansion order of the effective gyromagnetic ratio $g_{S_*}^{\rm 1SF}$ from the $O(u^{7.5})$ level given in Eq. (6.37) of \cite{Bini:2014ica} to the $O(u^{9.5})$ level. We list below the final result, expressed in the effective one-body radial variable $u$.

\begin{widetext}
\begin{eqnarray} 
g_{S_*}^{\rm 1SF}(u)&=& -\frac{3}{4} u-\frac{39}{4} u^2+\left(\frac{41}{32}\pi^2-\frac{7627}{192}\right) u^3\nonumber\\
&+&\left(-48\gamma+\frac{23663}{2048}\pi^2-\frac{1456}{15}\ln(2)-\frac{1017}{20}-24\ln(u)\right) u^4\nonumber\\
&+&\left(-\frac{729}{7}\ln(3)+\frac{9832}{35}\gamma+\frac{712905}{4096}\pi^2+\frac{70696}{105}\ln(2)-\frac{161160813}{89600}+\frac{4916}{35}\ln(u)\right) u^5\nonumber\\
&-&\frac{93304}{1575}\pi  u^{11/2} \nonumber\\
&+& \left(\frac{315657}{280}\ln(3)+\frac{480829}{2835}\gamma+\frac{16790137}{1048576}\pi^4-\frac{674904611}{7077888}\pi^2-\frac{2954531}{2835}\ln(2)-\frac{18167439833}{7257600}+\frac{480829}{5670}\ln(u)\right) u^6\nonumber\\
&+&\frac{4596019}{12600}\pi u^{13/2}\nonumber\\
&+& \left(-\frac{12227517}{3080}\ln(3)-1088\zeta(3)-\frac{1953125}{3564}\ln(5)-\frac{903605468}{121275}\gamma+\frac{58208}{105}\gamma^2-\frac{204902966117}{335544320}\pi^4\right. \nonumber\\
&&+\frac{1167584}{525}\ln(2)^2-\frac{7532631301}{9175040}\pi^2+\frac{499904}{225}\ln(2)\gamma-\frac{5587843424}{779625}\ln(2)+\frac{48146264595158227}{625895424000}\nonumber\\
&& \left.-\frac{451802734}{121275}\ln(u) +\frac{58208}{105}\gamma\ln(u)+\frac{249952}{225}\ln(2)\ln(u)+\frac{14552}{105}\ln(u)^2\right) u^7\nonumber\\
&&+\frac{118299749}{2182950}\pi  u^{15/2}\nonumber\\
&+&\left(-\frac{52964727700527}{3139136000}\ln(3)+\frac{141648}{35}\zeta(3)+\frac{366384765625}{41513472}\ln(5)+\frac{4204284206047}{264864600}\gamma-\frac{10974904}{3675}\gamma^2\right. \nonumber\\
&&-\frac{1135089788764019}{42949672960}\pi^4-\frac{20022888}{1225}\ln(2)^2+\frac{142155}{49}\gamma\ln(3)+\frac{142155}{49}\ln(2)\ln(3)+\frac{1241427590810221}{369937612800}\pi^2\nonumber\\
&& -\frac{164036944}{11025}\ln(2)\gamma+\frac{90305230479881}{3972969000}\ln(2)+\frac{142155}{98}\ln(3)^2+\frac{1094977266529990589159}{427173626880000}\nonumber\\
&&\left. +\frac{4189028005087}{529729200}\ln(u)-\frac{10974904}{3675}\gamma\ln(u)-\frac{82018472}{11025}\ln(2)\ln(u)+\frac{142155}{98}\ln(u)\ln(3)-\frac{2743726}{3675}\ln(u)^2\right) u^8\nonumber\\
&&+\left(\frac{142517152}{55125}\ln(2)\pi-\frac{6965217870900563}{762810048000}\pi+\frac{213592544}{165375}\pi\gamma-\frac{1996192}{4725}\pi^3+\frac{106796272}{165375}\pi\ln(u)\right) u^{17/2}\nonumber\\
&+&\left(\frac{15312301495292259}{69060992000}\ln(3)+\frac{1294640}{81}\zeta(3)-\frac{13759767578125}{249080832}\ln(5)+\frac{16569352454284793}{202280878800}\gamma\right. \nonumber\\
&&-\frac{4477353976}{1403325}\gamma^2-\frac{691974898583334793}{5497558138880}\pi^4+\frac{540743945464}{9823275}\ln(2)^2-\frac{847418031}{26950}\gamma\ln(3)-\frac{847418031}{26950}\ln(2)\ln(3)\nonumber\\
&&-\frac{1368675122796890401}{146495294668800}\pi^2+\frac{264895105264}{9823275}\ln(2)\gamma-\frac{207604582525}{402653184}\pi^6-\frac{3003559322617}{1111968000}\ln(7)\nonumber\\
&&-\frac{78318056677502249}{7079830758000}\ln(2)-\frac{847418031}{53900}\ln(3)^2+\frac{849725095980588589949507303}{66987659472814080000}+\frac{16550580788732153}{404561757600}\ln(u)\nonumber\\
&&\left. -\frac{4477353976}{1403325}\gamma\ln(u)+\frac{132447552632}{9823275}\ln(2)\ln(u)-\frac{847418031}{53900}\ln(u)\ln(3)-\frac{1119338494}{1403325}\ln(u)^2\right) u^9\nonumber\\
&&+\left(-\frac{78150479}{4410}\ln(2)\pi+\frac{298035034972802327}{11783531760000}\pi-\frac{1137638861}{154350}\pi\gamma+\frac{4065633}{1372}\pi\ln(3)+\frac{23679911}{13230}\pi^3\right. \nonumber\\
&& \left. -\frac{1137638861}{308700}\pi\ln(u)\right) u^{19/2}+O_{\ln{}}(u^{10})\,.
\end{eqnarray}
\end{widetext}

Here the coefficients of $u^6$ and $u^7$ differ from the ones given in Ref. \cite{Bini:2014ica}  because of the corresponding change  in $\delta\psi$.
More precisely,  the terms to be added to Eqs. (6.36)$_1$ and (6.37)$_1$ in \cite{Bini:2014ica}  read 
\beq
\Delta c_6^{g_{S*}}=\Delta c_7^{\delta\psi}
\eeq
and
\beq
\Delta c_7^{g_{S*}}=\Delta c_8^{\delta\psi}-\frac32 \Delta c_7^{\delta\psi}\,.
\eeq
Let us also point out a misprint in the expression of $g_7^{\ln{}}$ given in Eq. (6.37)$_2$ of  \cite{Bini:2014ica}:  
the additional term
\beq
+\frac{249952}{225}\ln 2\,,
\eeq
was inadvertently omitted.

\noindent {\bf Acknowledgments}  

We thank 
Sam  Dolan, Patrick Nolan, Adrian  Ottewill, Niels Warburton, Barry Wardell and Chris Kavanagh for useful discussions, exchange of data and constructive criticism.
We also thank Abhay Shah  for informative  email exchanges.
Finally we are grateful to ICRANet for partial support. 
D.B. thanks the Italian INFN (Naples) for partial support and IHES for hospitality during the development of this project.

\end{document}